\documentclass[preprint,showpacs,preprintnumbers,amsmath,amssymb]{revtex4-2}

\usepackage{graphicx}
\usepackage{dcolumn}
\usepackage{bm}
\usepackage{color}

\begin{document}

\newif\ifplot
\plottrue
\newcommand{\RR}[1]{[#1]}
\newcommand{\intsum}{\sum \kern -15pt \int}
\newfont{\Yfont}{cmti10 scaled 2074}
\newcommand{\Y}{\hbox{{\Yfont y}\phantom.}}
\def\O{{\cal O}}
\newcommand{\bra}[1]{\left< #1 \right| }
\newcommand{\braa}[1]{\left. \left< #1 \right| \right| }
\def\Bra#1#2{{\mbox{\vphantom{$\left< #2 \right|$}}}_{#1}
\kern -2.5pt \left< #2 \right| }
\def\Braa#1#2{{\mbox{\vphantom{$\left< #2 \right|$}}}_{#1}
\kern -2.5pt \left. \left< #2 \right| \right| }
\newcommand{\ket}[1]{\left| #1 \right> }
\newcommand{\kett}[1]{\left| \left| #1 \right> \right.}
\newcommand{\scal}[2]{\left< #1 \left| \mbox{\vphantom{$\left< #1 #2 \right|$}}
\right. #2 \right> }
\def\Scal#1#2#3{{\mbox{\vphantom{$\left<#2#3\right|$}}}_{#1}
{\left< #2 \left| \mbox{\vphantom{$\left<#2#3\right|$}}
\right. #3 \right> }}

\title{Three-nucleon force
  effects in  inclusive spectra of the neutron-deuteron breakup reaction}

\author{H.\ Wita{\l}a$^{1}$}
\email{henryk.witala@uj.edu.pl}
\author{J.\ Golak$^{1}$}
\author{R.\ Skibi\'nski$^{1}$}
\author{V.\ Soloviov$^{1}$}
\author{K.\ Topolnicki$^{1}$}
\affiliation{
$^{1}$M. Smoluchowski Institute of Physics, Jagiellonian University,
PL-30348 Krak\'ow, Poland}

\date{\today}

\begin{abstract}
  We investigate the sensitivity of the non-exclusive nucleon induced
  deuteron breakup reaction to the three-nucleon interaction 
and distributions of
  three-nucleon force effects in inclusive spectra.
  To this end we solve the three-nucleon Faddeev equation at a number
  of incoming nucleon laboratory energies 
  using the CD~Bonn nucleon-nucleon interaction alone or
  combined with the 2$\pi$-exchange Tucson-Melbourne three-nucleon force.
  Based on these solutions energy spectra of an outgoing nucleon, at
  a specified detection angle as well as spectra integrated over that angle,
  are calculated. By integrating the spectra at a given angle
  over the energy of the outgoing nucleon the angular
  distributions of three-nucleon force effects in the breakup process are 
  additionally obtained. 
  Contrary to elastic nucleon-deuteron scattering, where at higher energies
  significant three-nucleon force effects were encountered for scattering angles around
  the minimum of the cross section, for the breakup process
  only moderate effects are found and they are restricted to forward angles.
Results of the present
  investigation show that the large three-nucleon force effects found for some
  specific complete  breakup configurations are reduced substantially in the
  incomplete spectra when averaging over contributing complete geometries
  is performed.
\end{abstract}


\maketitle

\section{Introduction}

Studies of the three-nucleon (3N) continuum revealed significant 
three-nucleon force (3NF) effects in the elastic nucleon-deuteron (Nd)
scattering and the deuteron breakup reactions. Namely for 
laboratory energies of the incoming nucleon above $\approx 60$~MeV  large 
discrepancies between theoretical predictions and data
 were found  in the angular distributions of the
elastic Nd  scattering observables 
 \cite{wit98,witelas2001,Rep.Prog.Phys.75.016301}
  as well as in the total cross section for neutron-deuteron (nd)
  scattering \cite{abfalt98,wittot99}.
  Also in some kinematically complete breakup configurations large
  changes of the cross section  caused by 3NF's were 
  predicted \cite{kuros_br}. Generally the detected 3NF effects  grow with
  the increasing energy of the incoming nucleon. The commonly used
  (semi)phenomenological long-range $2\pi$-exchange 3NF's, such as
the Tucson-Melbourne (TM) \cite{TM99} or the Urbana IX \cite{uIX}  
  when combined
  with (semi)phenomenological, high-precision  nucleon-nucleon (NN) potentials,
  such as the AV18 \cite{AV18}, the CD~Bonn \cite{cdb}, the Nijm1 or the Nijm2 \cite{nijm} forces
are able to explain the dominant part of the discrepancy 
  for the incoming nucleon energies up to about 135~MeV. 
   However, at still higher 
  energies a significant part of the deviation from data remains unexplained
  \cite{witelas2001,wittot99}. 
  Since the relativistic 3N Faddeev calculations \cite{witrel1,witrel2} 
  showed only negligible effects of relativity for the elastic scattering 
  observables and the
  total nd cross section at energies up to the $\pi$ production
  threshold, therefore those remaining discrepancies indicate the action
  of short-range components of the 3NF absent in the above mentioned models. 

  Since the total cross section for nd scattering is a sum of the total
  elastic scattering and breakup cross sections, 
  interesting questions arise 
  about importance of 3NF effects in incomplete spectra
  of the breakup process as well as on their distribution and
  dependence on the  incoming neutron energy.  
  To answer them we investigated,
at a number of incoming neutron energies, in the range 14-294~MeV,
  the energy spectra of the outgoing nucleon, taken as a proton or a neutron,
  in the incomplete breakup. 
  We examined energy spectra at
  a specific laboratory angle of the outgoing nucleon as well as the energy
  spectra arising from integrations over this angle. 
  Integrating the energy spectra at a specific angle of the outgoing
  nucleon over the allowed nucleon energy enabled us to determine
  angular distributions of the 
  single-nucleon inclusive breakup cross sections.
  It permitted us to determine how 3NF effects are spread 
  over the angular domain in incomplete breakup 
  and to compare it to the angular distribution of 3NF effects in
  the elastic nd scattering.

The paper is organized as follows: in Sec. \ref{form} we 
describe briefly the underlying theoretical formalism leading to predictions
for different energy spectra. 
 We present and discuss our results  in 
 Sec. \ref{results}. Finally, we summarize and 
conclude in Sec. \ref{summary}.

 \section{Single-nucleon energy spectra and angular 
distributions in the 
   deuteron breakup reaction}
\label{form}

Neutron-deuteron scattering with the nucleons interacting
through a nucleon-nucleon potential $v_{NN}$ and a three-nucleon force
 $V_4 = V_4^{(1)} + V_4^{(2)} + V_4^{(3)}$ is
described in terms of the breakup operator $T$ satisfying the
Faddeev-type integral equation~\cite{wit88,glo96,hub97,book}
\begin{eqnarray}
T \vert \phi \rangle  &=& \ t \, P \, \vert \phi \rangle  \ 
+ \ ( 1 + t G_0 ) \, V_4^{(1)} \, ( 1 + P ) \, \vert \phi \rangle  \
+  \ t \, P \, G_0 \, T \vert \phi \rangle \ \cr
&+& \ ( 1 + t G_0 ) \, V_4^{(1)} \, ( 1 + P ) \, G_0 \, T \vert \phi \rangle ~,
\label{eq1a}
\end{eqnarray}
The two-nucleon $t$-matrix $t$ is the solution of the
Lippmann-Schwinger equation with the interaction
$v_{NN}$.  The permutation operator $P=P_{12}P_{23} +
P_{13}P_{23}$ is given in terms of the transposition operators,
$P_{ij}$, which interchange nucleons i and j.  The incoming state $
\vert \phi \rangle = \vert \vec {q}_0 \rangle \vert \varphi_d \rangle $
describes the free nucleon-deuteron motion with relative momentum
$\vec {q}_0$ and the deuteron wave function $\vert \varphi_d \rangle$.
Finally, $G_0$ is the resolvent of the three-body center of mass kinetic
energy. Each $V_4^{(i)}$ part of 3NF is symmetric under the 
exchange of the nucleons
$j$ and $k$ ($i, j, k=1,2,3$ and $j\ne i \ne k \ne j$). 

The transition amplitudes for the elastic nd scattering,
$ \langle \phi' \vert U \vert \phi \rangle $, and
breakup reactions, 
$  \langle \phi_0 \vert U_0 \vert \phi \rangle $, 
are given in terms of $T$ by~\cite{wit88,glo96,hub97,book}:
\begin{eqnarray}
  \langle \phi' \vert    U \vert \phi \rangle &=& \langle \phi' \vert
  P G_0^{-1} \ + \ V_4^{(1)} \,
  ( 1 + P ) \ + \ 
\ P T \  + \
V_4^{(1)} \, ( 1 + P ) \, G_0 \, T \vert \phi \rangle ~, \cr
\langle \phi_0  \vert U_0 \vert  \phi \rangle &=&  \langle \phi_0  \vert
(1+P)T \vert \phi \rangle ~.
\label{eq1c}
\end{eqnarray}
In the latter case the transition amplitude comprises 
a final breakup state
 $ \vert \phi_0 \rangle = \vert \vec p_1 \vec q_1 m_1 m_2 m_3\rangle$ of
three outgoing nucleons 
defined by individual nucleon spin projections $m_i$ and 
by two relative Jacobi momenta
$\vec p_1$ and $\vec q_1$, which are linear combinations
of the individual nucleon momenta $\vec k_i$:
\begin{eqnarray}
 \vec p_i &=& \frac {1} {2} (\vec k_j - \vec k_k ) \cr
 \vec q_i &=& \frac {2} {3} [ \vec k_i - \frac{1} {2} (\vec k_j + \vec k_k ) ]~,
\label{eq1d}
\end{eqnarray}
for $\{ i,j,k \} = \{ 1,2,3 \} $ and cyclic permutations. 
The center of mass energy of the 3N system
 $E_{c.m.}$ is specified by incoming
relative nucleon-deuteron momentum $\vec q_0$ and the deuteron binding energy
$E_d$:
\begin{eqnarray}
  E_{c.m.} = \frac {3} {4m} q_0^2  + E_d \equiv \frac {3} {4m} q_{max}^2
  =  \frac {3} {4m} q_i^2 + \frac {1} {m} p_i^2 ~,
\label{eq1e}
\end{eqnarray}
where $m$ is the nucleon mass.

\begin{figure}[htbp] 
\includegraphics[scale=0.9]{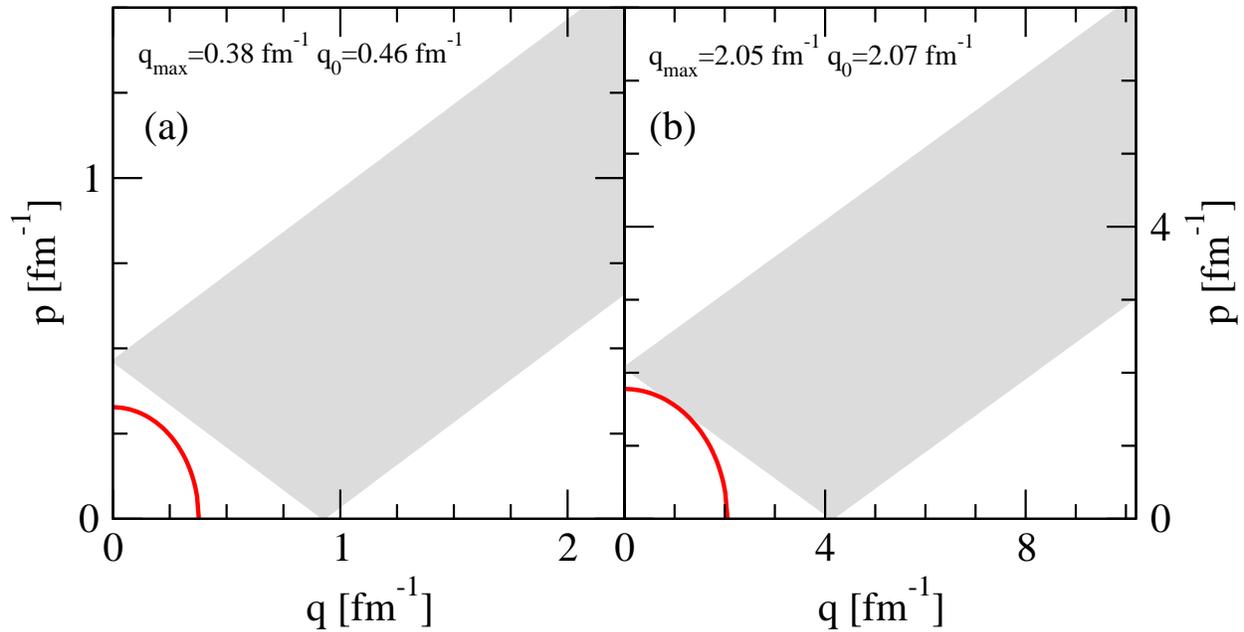}
\caption{(Color online) 
  Regions in the  ($q-p$) plane (see text) 
which contribute to the breakup
  reaction ((red) solid line) and 
to the 
  $ \langle \phi' \vert PT \vert \phi \rangle$ term
of the elastic scattering amplitude
  (gray highlighted region) at the 
  incoming nucleon laboratory energy $E=10$ (a) and $200$~MeV (b).
\label{fig1}}
\end{figure}

 It follows from Eq.~(\ref{eq1c}) that contributions 
 to a particular kinematically complete breakup configuration, 
 specified by  momenta of three outgoing nucleons,  
 are given by three matrix elements 
 $\langle \vec p_i  \vec q_i m_1 m_2 m_3 \vert T \vert \phi \rangle $ determined at 
 three pairs of momentum magnitudes ($p_i,q_i$) 
 lying on an ellipse in the ($q-p$) plane, 
 given by Eq.~(\ref{eq1e}) (see Fig.\ref{fig1}).
 Performing exclusive or inclusive breakup measurements one
 is restricted to points lying on that ellipse. 
 While the exclusive breakup is 
 very selective, being restricted to only three ($p_i,q_i$) points,
 in the incomplete breakup one integrates over
 contributing complete geometries along that curve.
Thus the incomplete breakup delivers information on the underlying dynamics averaged over
configurations which are taken into account.
 In contrast to the breakup reaction, the elastic Nd scattering
 receives contributions from
 practically all regions of the ($q-p$) plane, due to the integration over the
  relative momentum of the two nucleons forming the deuteron. 
  It is interesting to note that the region of the ($q-p$) plane which
  contributes to the
 elastic scattering transition amplitude by the dominant  
 $ \langle \phi' \vert PT \vert \phi \rangle$  term does not
 overlap with the ellipse of contributions to the breakup reaction. Namely,
 contributions of that term to elastic nd scattering
 come from the region of ($q,p$) values
 with $p \in ( \vert q_0 - \frac {1} {2} q \vert ,  q_0 + \frac {1} {2} q )$
 (see Appendix \ref{app_a}). In Fig. \ref{fig1} we exemplify that separation of
 breakup and elastic scattering regions of the ($q-p$) plane
for two laboratory energies of the incoming neutron:
 $E=10$ and $200$~MeV.
 
It follows that the sensitivity of breakup observables to the underlying dynamics,
   in particular their sensitivity to
 3NF effects, will  be different from the sensitivity of the elastic
 Nd scattering observables.
 Also the  averaging over many
 contributing kinematically complete geometries, should  reduce sensitivity of
 the incomplete breakup to the underlying dynamics.

 The reduction of sensitivity will depend on the complete
 configurations over which the averaging is done. Performing a
 standard incomplete  breakup measurement one outgoing nucleon is
 detected at a specific laboratory angle  and its energy spectrum is measured.
 In other conceivable incomplete measurements the energy spectrum of
 the outgoing nucleon stemming from some angular range can be determined. 
  Both types of
 energy spectra can be predicted theoretically  by performing proper
 integrations over the Jacobi momenta of the contributing complete
 configurations. Integration of the energy spectra at a specific angle over
 an energy of the outgoing nucleon provides the angular distribution of the
 incomplete breakup.

In order to investigate the sensitivity of such various spectra to the 3NF and
 to compare them with elastic nd scattering 
   we solved the 3$N$ Faddeev equation in a partial wave 
momentum-space basis for a number of the incoming nucleon laboratory 
energies $E=14, 70, 100, 135, 200, 250$ and $294$~MeV. As a NN interaction 
we used the high precision semi-phenomenological CD~Bonn potential \cite{cdb}.  
 We took   that interaction alone 
 or together with the TM99 3NF \cite{TM99} whose 
 cut-off parameter $\Lambda$ was 
adjusted  so that 
this particular combination of a NN- 
and 3N-force reproduced the experimental triton 
binding energy~\cite{witelas2001}.
When solving the 3N Faddeev equation  we  included all 3N partial
wave states   with the total two-nucleon angular momentum $j \le 5$ 
and the total 3N angular momentum   $J \le 25/2$.

\section{Results and discussion}
\label{results}

\begin{figure}[htbp] 
 \begin{center}
 \includegraphics[scale=0.4,angle=0]{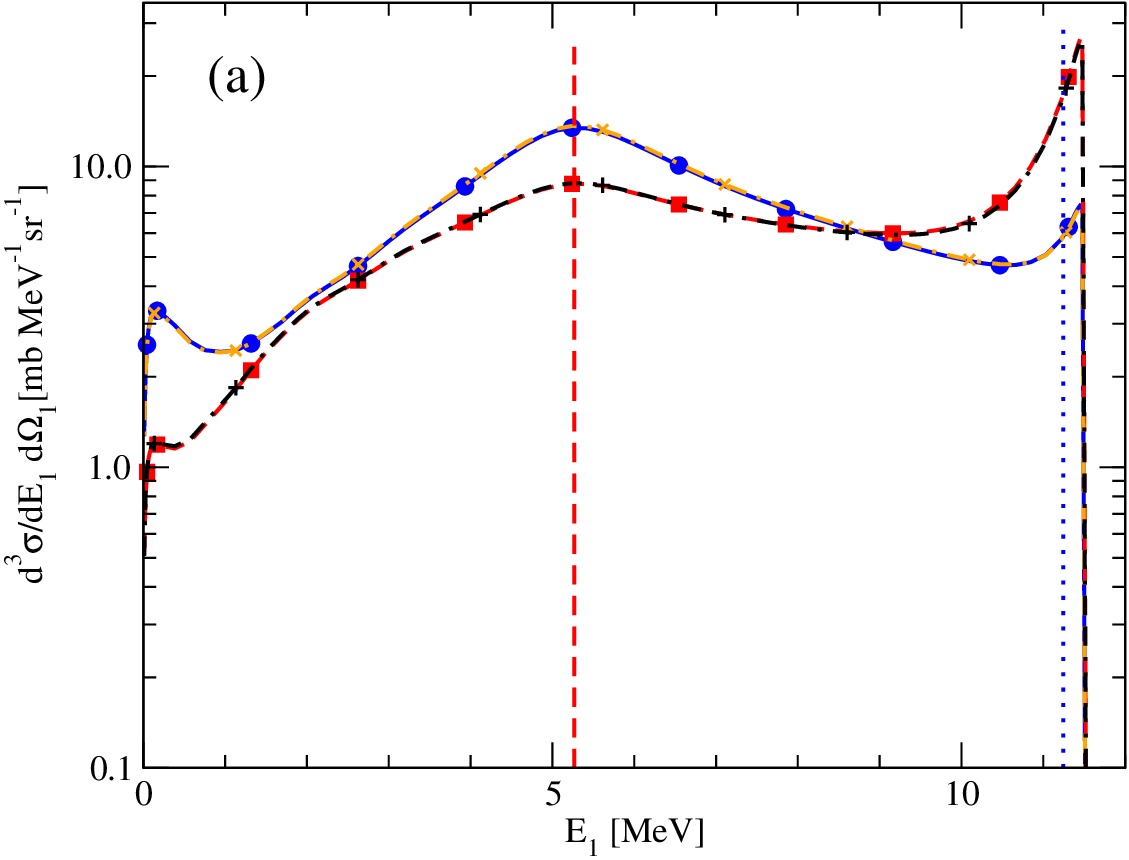}
\hspace{1em}    
\includegraphics[scale=0.4,angle=0]{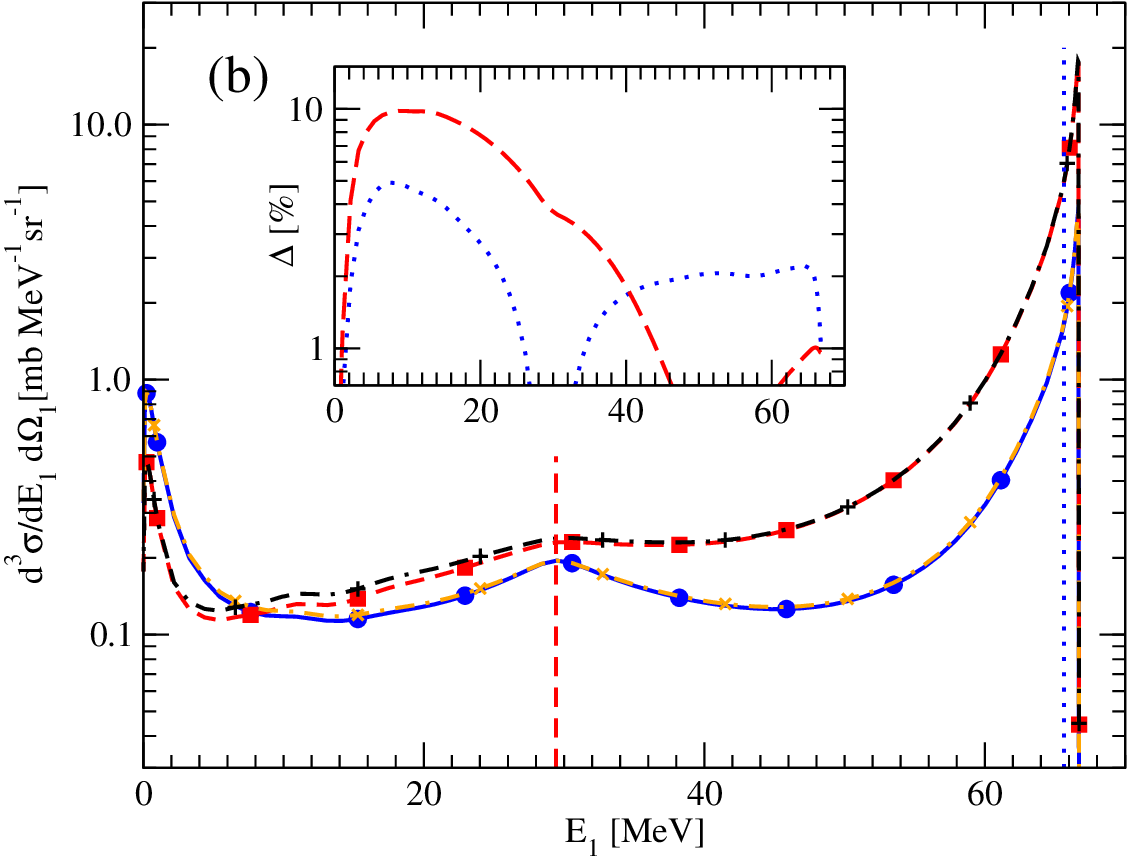}
 \end{center}
 \vspace{1em}
  \begin{center}
\includegraphics[scale=0.4,angle=0]{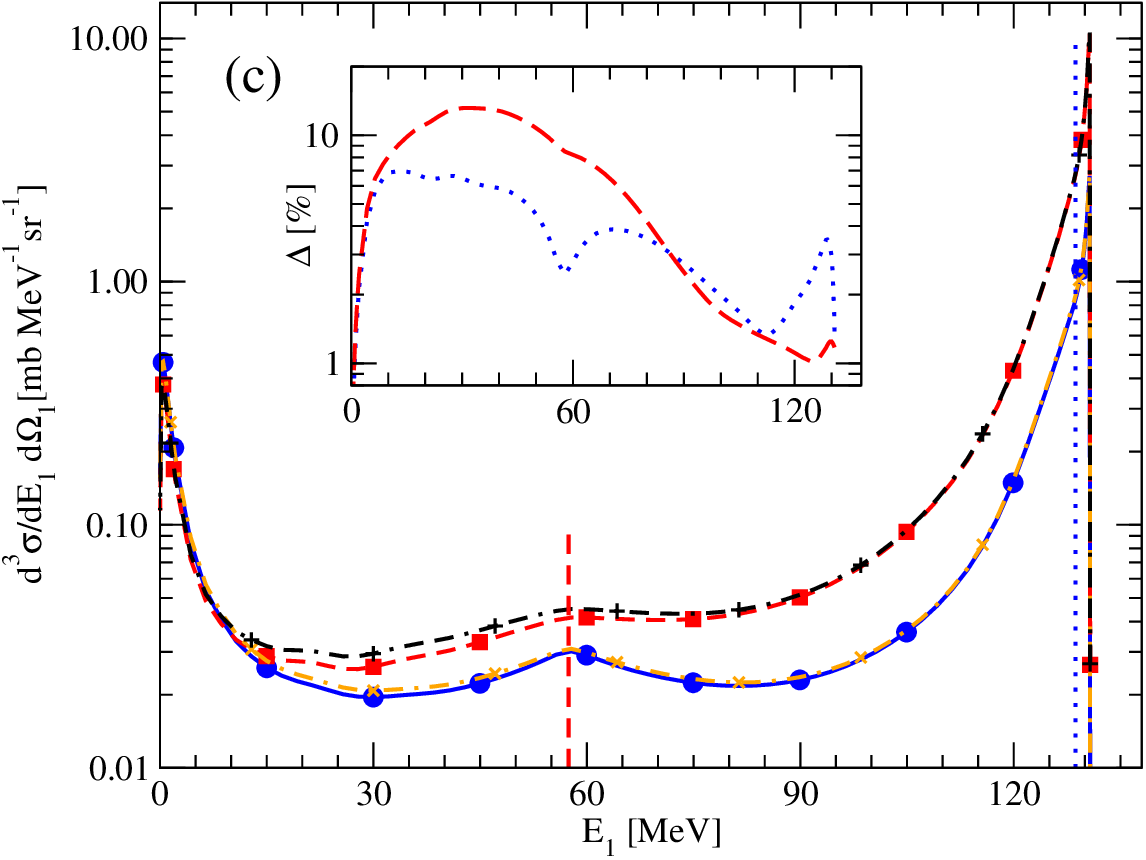}
\hspace{1em}
\includegraphics[scale=0.4,angle=0]{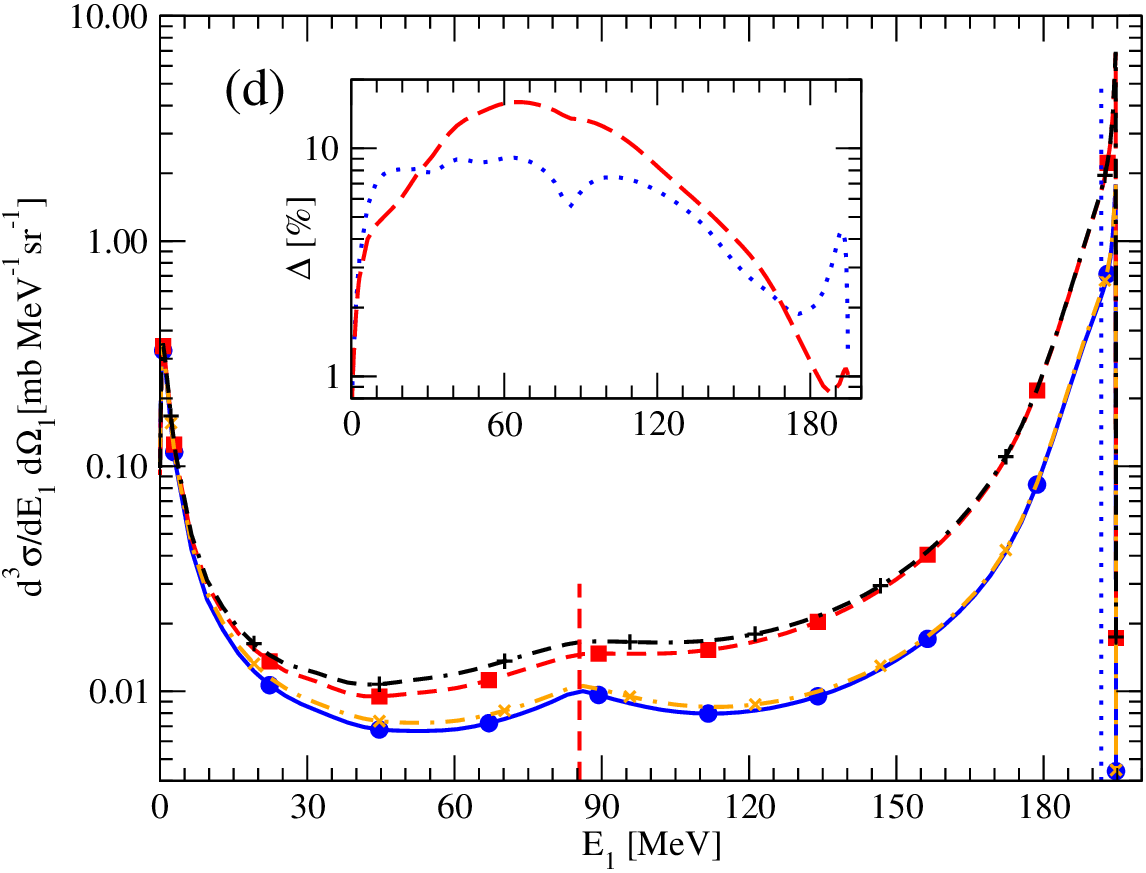}
 \end{center}
\caption{(Color online) 
  The threefold differential cross section $\frac {d^3\sigma} {d\Omega_1 dE_1}$
  of the outgoing nucleon detected at the 
  laboratory angle $\theta_1^{lab}=10^{\circ}$
  in the $d(n,N_1)N_2N_3$ breakup reaction for the laboratory
  energy of the incoming neutron $E$=14 (a), 70 (b), 135 (c), and
  200 (d)~MeV. The spectra of the outgoing neutron in d(n,n)np reaction are
  shown with the (blue) solid line, marked with circles when nucleons interact
  with the CD~Bonn potential only. Combining the CD~Bonn with the TM99 3NF gives the
  (orange) dashed-dotted line, marked with x-es. The spectra of the outgoing
  proton are given by the (red) dashed line, marked with squares,
  (the CD~Bonn only) and the (black) double-dashed-dotted line, marked with pluses,
  (the CD~Bonn +TM99). 
  The position of QFS(1-2) and QFS(1-3) is given by the (blue) dotted
  vertical line. The position of FSI(1-2) and FSI(1-3) is given by the (red) dashed
  vertical line.
  In inserts the magnitudes of 3NF effects
  defined by $\Delta \equiv \frac {\sigma(NN+3NF)-\sigma(NN)} {\sigma(NN)}
  \times 100 \%$ are shown with the (blue)
  dotted  line for the outgoing neutron and with the (red) long-dashed line for the outgoing
  proton.
\label{fig2}}
\end{figure}

\begin{figure}[htbp] 
 \begin{center}
 \includegraphics[scale=0.4,angle=0]{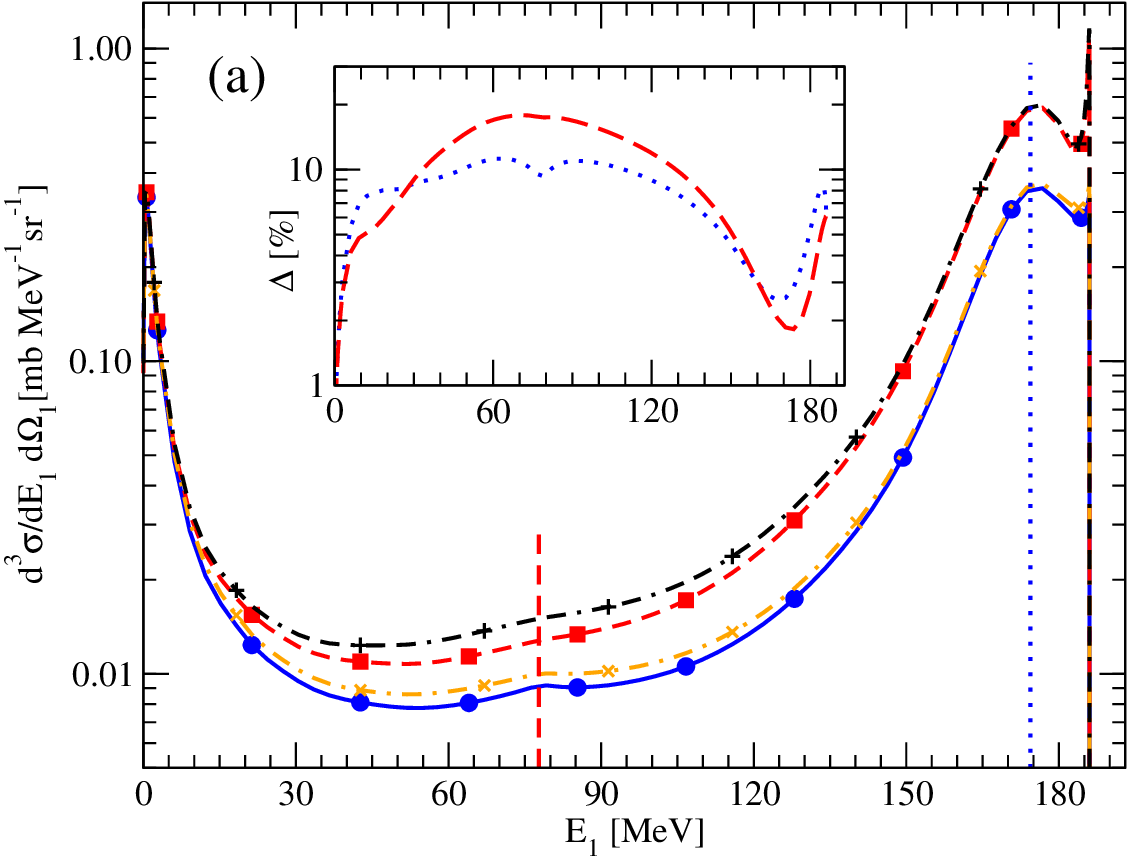}
\hspace{1em}    
\includegraphics[scale=0.4,angle=0]{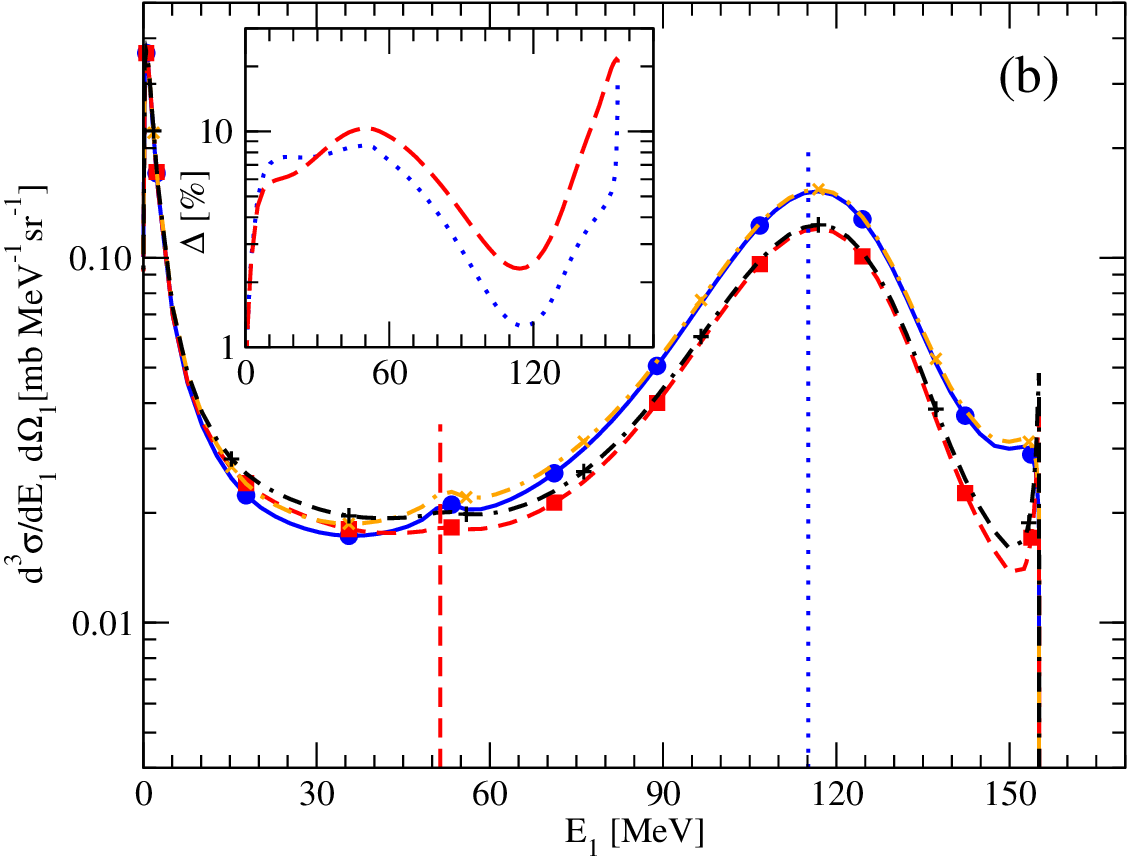}
 \end{center}
 \vspace{1em}
  \begin{center}
\includegraphics[scale=0.4,angle=0]{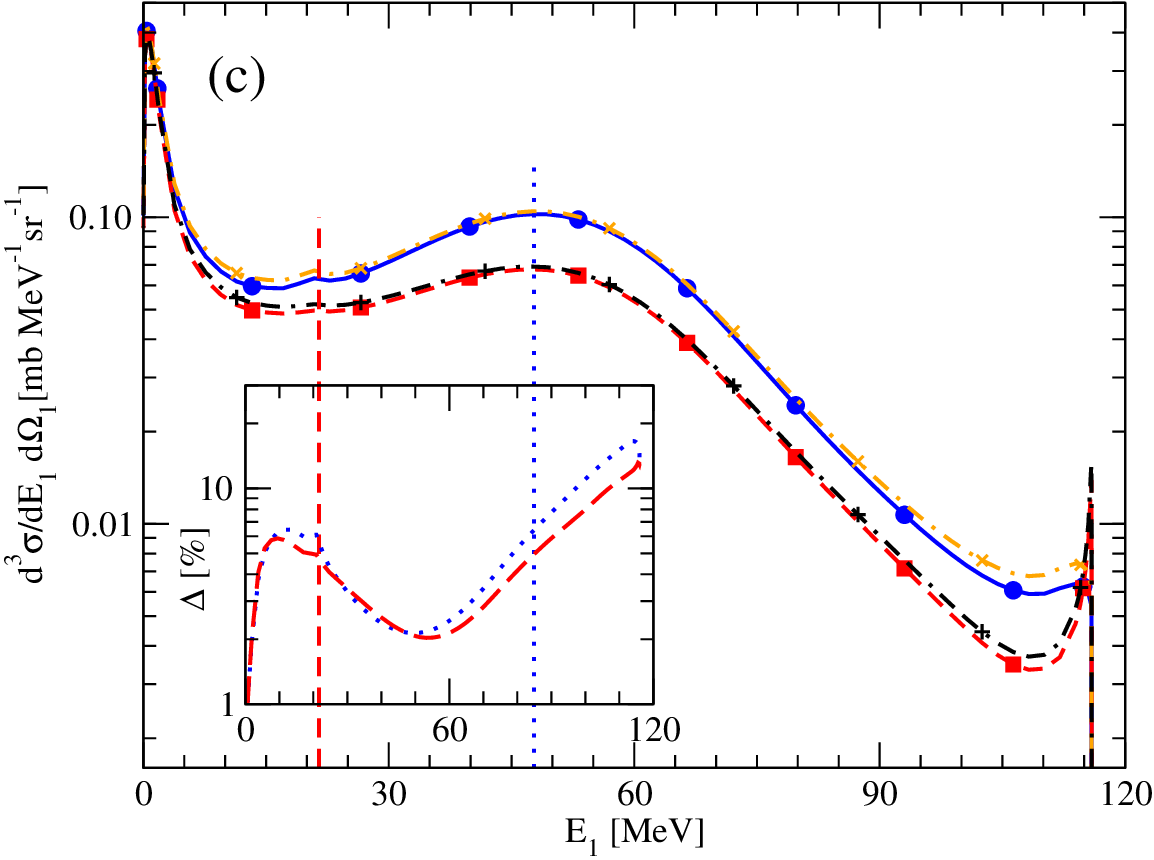}
\hspace{1em}
\includegraphics[scale=0.4,angle=0]{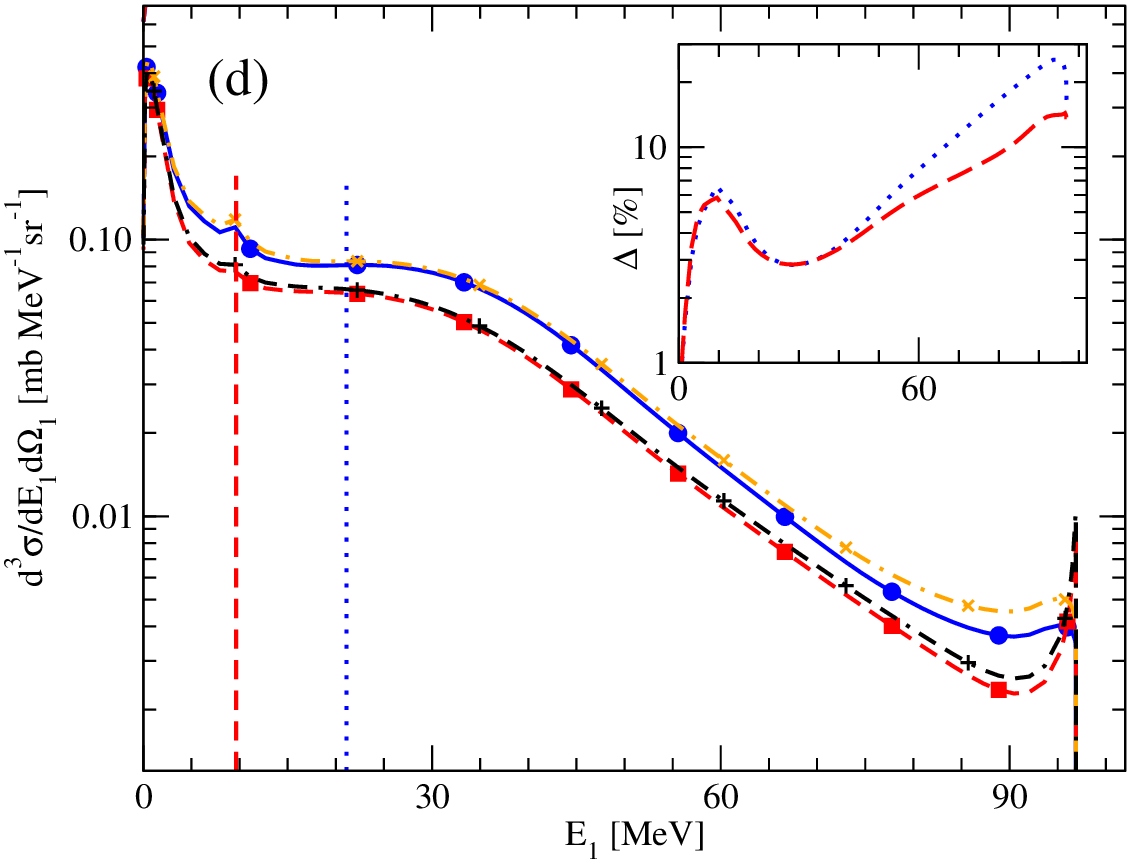}
 \end{center}
\caption{(Color online) 
  The energy spectra $\frac {d^3\sigma} {d\Omega_1 dE_1}$
  of the outgoing nucleon 
  at the laboratory angles $\theta_1^{lab}=20^{\circ} (a), 40^{\circ} (b), 60^{\circ}$ (c), and $70^{\circ} (d)$
  in the $d(n,N_1)N_2N_3$ breakup reaction for the laboratory
  energy of the incoming neutron $E=200$~MeV. For description of lines
 and inserts  see Fig. \ref{fig2}. 
\label{fig2a}}
\end{figure}

In Fig.~\ref{fig2} we show the threefold differential cross section $\frac{d^3\sigma}{d\Omega_1dE_1}$
as a function of the energy $E_1$. Specifically, in Fig.~\ref{fig2} the energy spectra of the outgoing nucleon
 (neutron or proton) from   incomplete breakup $d(n,N_1)N_2N_3$, detected at
a laboratory angle $\theta_1^{lab}=10^{\circ}$ are exemplified at four incoming neutron laboratory
energies $E=14, 70, 135$, and $200$~MeV.
 These spectra unveil a characteristic
 structure, with a peak at the highest energy of the outgoing nucleon,
 which is due to
 a strong final state interaction of nucleons $2$ and $3$ (FSI(2-3)),
 having a small relative energy, 
in a nucleon-nucleon partial wave state $^1S_0$. 
The enhancement of the cross section in the region of vanishing energy
of nucleon 1 results from approaching the quasi-free-scattering (QFS)
complete breakup geometry defined by the condition ${\vec p}_1 = 0$.
 That QFS configuration corresponds  to quasi free scattering of nucleons $2$
 and $3$ (QFS(2-3)). 
 In the spectra one can see also two additional dominating contributions coming
  from specific kinematically complete breakup configurations
 with large cross sections, whose positions
 are indicated by vertical dotted and dashed lines.
  One of them is QFS(1-2) ($\vec p_3=0$)
  and QFS(1-3) ($\vec p_2=0$) occurring at $E_1$ indicated by the (blue) dotted
  vertical line, 
  and appearing in the energy spectrum at laboratory angles of nucleon $1$
  below some angle $\theta_1^{max}$ given  
 in terms of the incoming neutron laboratory energy $E$ and the deuteron
 binding energy $E_d$ by $\theta_1^{max}=\arcsin(\sqrt{-\frac {2E_d}  {E} })$.
 The second dominating contribution comes from the kinematically complete breakup
 geometry FSI(1-2) (kinematical condition $\vec p_1=\vec p_2$)
 and FSI(1-3) ($\vec p_1=\vec p_3$), which  occur at $E_1$ indicated by
  a (red) vertical dashed line, 
 and appears in the energy spectrum up to a laboratory angle
 $\theta_1^{max}=\arcsin(\sqrt{-\frac {3E_d}  {2E} })$. 
 These positions move when changing the angle of the outgoing nucleon, what
 is exemplified   in Fig.~\ref{fig2a} at the incoming  energy $E=200$~MeV and 
  four angles $\theta_1^{lab}=20^{\circ}, 40^{\circ}, 60^{\circ}$, and $70^{\circ}$.

While the structure of the threefold differential cross section for the outgoing neutron at given $\theta_1$
is similar to that of the outgoing  proton, the behaviour of the incomplete cross section
with respect to the energy and detection angle dependes on the isospin projection of the detected particle.
For example at $E=200$~MeV (see Figs.~\ref{fig2}
  and \ref{fig2a}) the neutron energy spectrum
 at forward angles lies below the
 proton one but with the increasing angle it overshoots
 the proton spectrum. The 3NF effects start to appear at $E=70$~MeV
 (see Fig.~\ref{fig2}), similarly
 as in the elastic Nd scattering process. 
 At forward angles ($10^{\circ}$) they are concentrated at the lower part of
 the spectrum where they enhance the cross section predicted with NN potential only. 
 At $70$~MeV they change the cross section by up to $10\%$, strengthening
 this effect to $\approx 20\%$ at $200$~MeV. 
  With the increasing detection angle the region of  
  significant changes of the cross section  extends to higher energies $E_1$. At
  larger angles and energies the magnitude of
  the cross section is however reduced by a factor of about 10.

\begin{figure}[htbp] 
   \begin{center}
     \includegraphics[scale=0.4,angle=0]{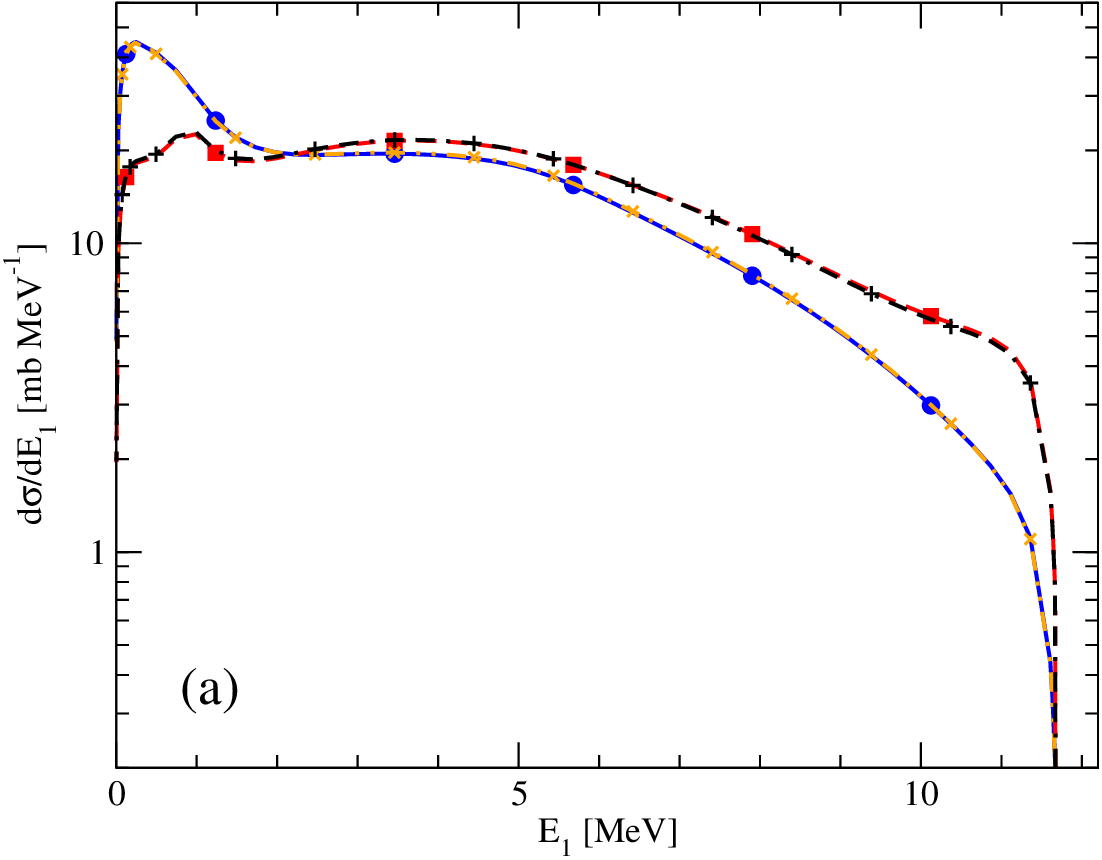}
  \hspace{1em}     
\includegraphics[scale=0.4,angle=0]{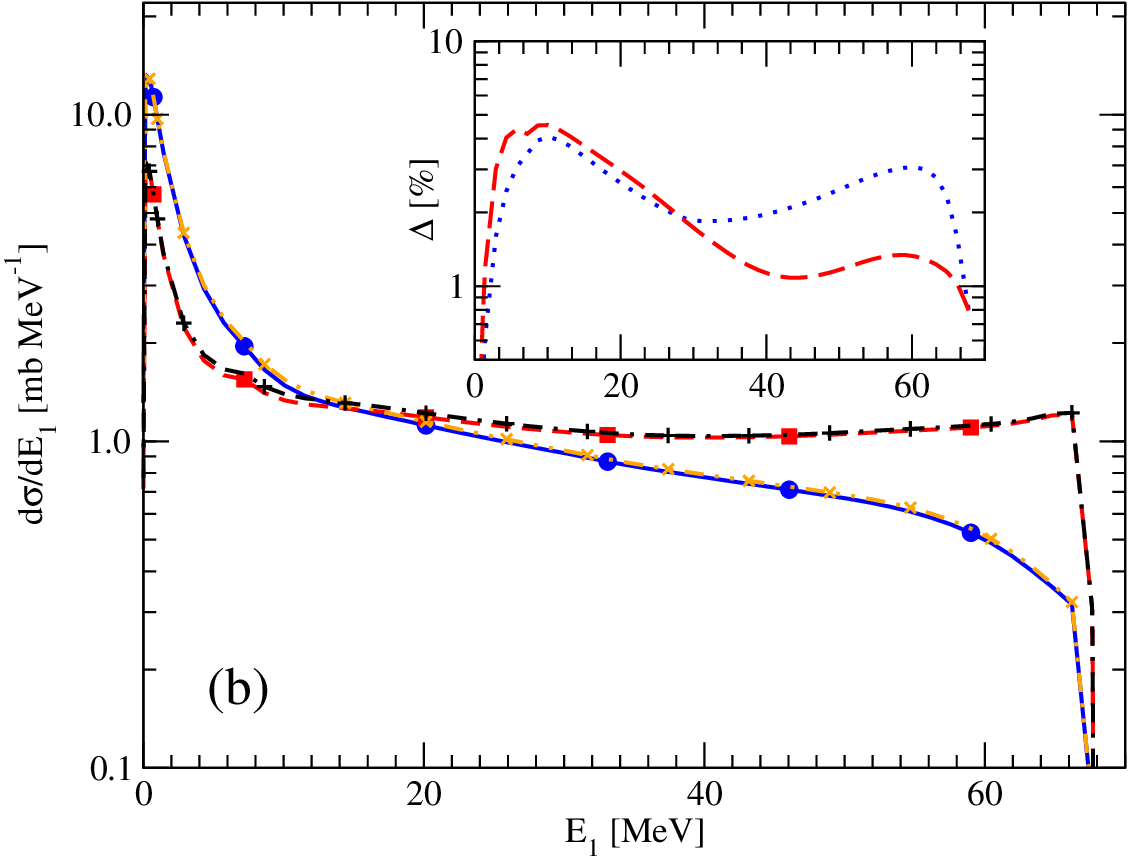}
   \end{center}
    \vspace{1em}
 \begin{center}
   \includegraphics[scale=0.4,angle=0]{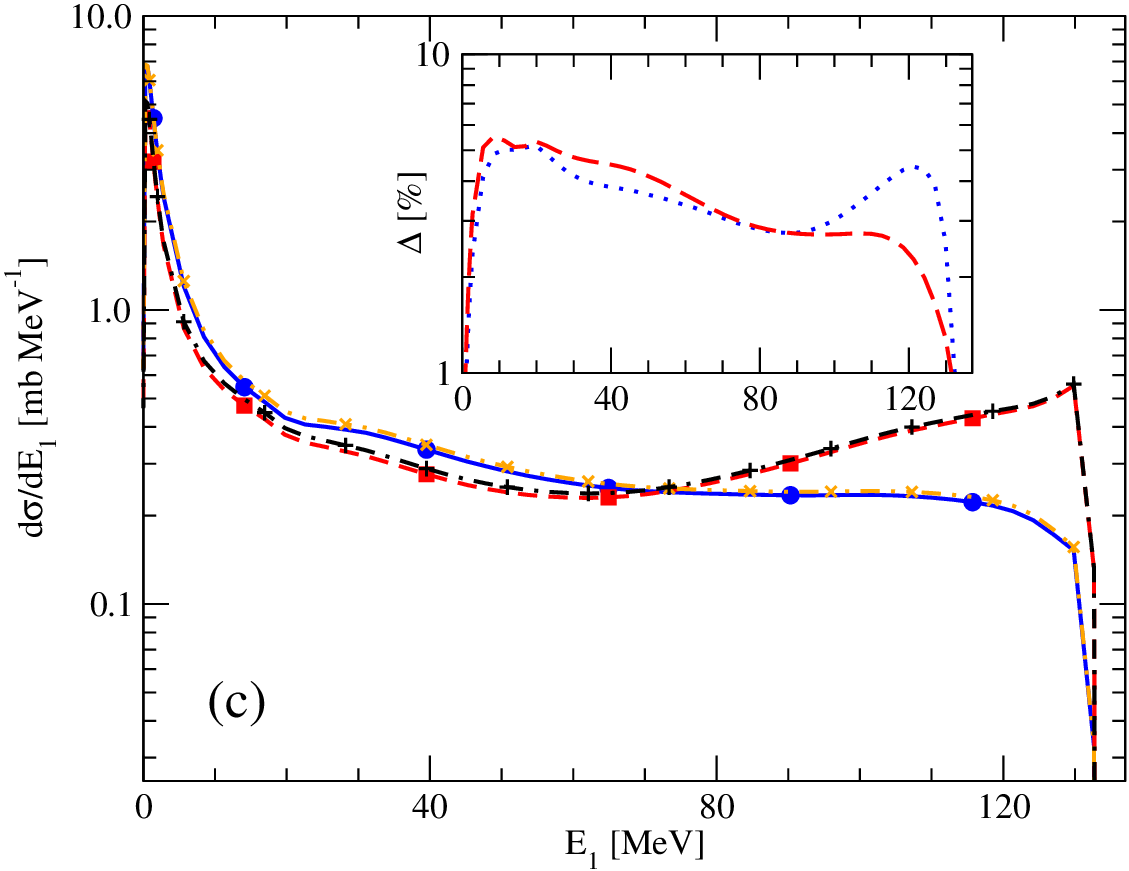}
   \hspace{1em}  
\includegraphics[scale=0.4,angle=0]{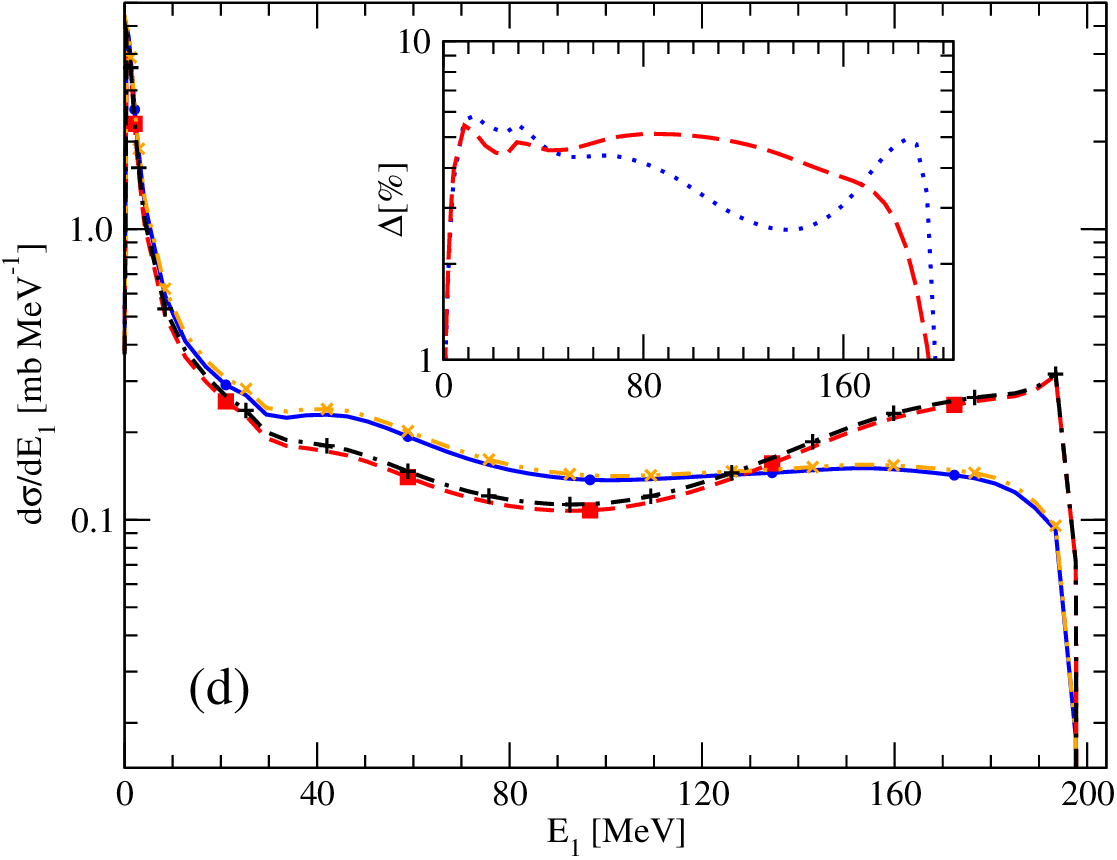}
 \end{center}
 \caption{(Color online)
The angle-integrated energy spectra $\frac {d\sigma} {dE_1}$
of the outgoing nucleon in the $d(n,N_1)N_2N_3$
 breakup reaction for the laboratory
  energy of the incoming neutron $E$=14 (a), 70 (b), 135 (c), and
  200 (d)~MeV. The spectra of the outgoing neutron in d(n,n)np reaction are
  given with the (blue) solid line, marked with circles when nucleons interact
  with the CD~Bonn potential only. Combining the CD~Bonn with the TM99 3NF gives
  the (orange) dashed-double-dotted line, marked with x-es.
  The spectra of the outgoing
  proton are given by the (red) dashed line, marked with squares,
  (the CD~Bonn only) and the (black) double-dashed-dotted line, marked with pluses,
  (the CD~Bonn+TM99). For explanation of inserts see Fig.~\ref{fig2}.
\label{fig3}}
\end{figure}

  This behavior of the cross section together with the 
  distributions of the 3NF effects
  are reflected in the angle-integrated energy spectra, $\frac{d\sigma}{dE_1}$, shown
  in Fig.~\ref{fig3}.
  The angular integration reduces contributions from
  the threefold differential cross section at forward angles
  with large 3NF effects, leading to the angle-integrated energy spectra
  with 3NF effects   distributed more or less
  uniformly along the spectrum with the magnitude of effects changing
  from $\approx 2-3\%$
at $70$~MeV to $\approx 4-6\%$ at $200$~MeV. A reduction of  3NF effects
when going from the spectra at given angle to angle integrated spectra
exemplifies a reduction of sensitivity of the incomplete breakup
due to averaging over contributing complete geometries.
While for complete breakup in specific configurations 3NF effects
of magnitude up to $\approx 90\%$ were found at $200$~MeV \cite{kuros_br},
 in the energy
spectra at a specific angle of the outgoing nucleon they diminish to
$\approx 20\%$ and reduce further to $\approx 6\%$ when additional
integration over the angle is performed. 
One can expect that further averaging by performing integrations over the
energy of the outgoing nucleon would lead to even smaller 3NF effects in
the total breakup cross section.
 
\begin{figure}[htbp] 
    \begin{center}
      \includegraphics[scale=0.4,angle=0]{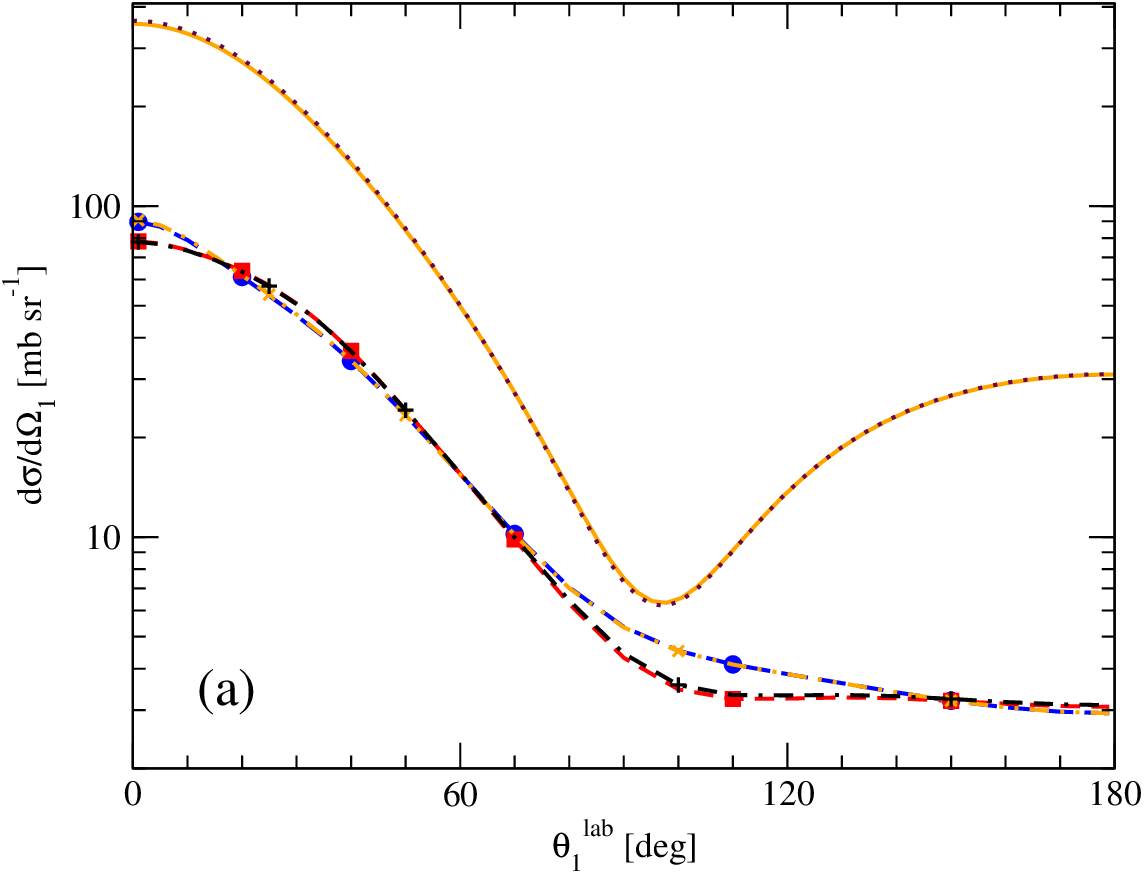}
      \hspace{1em} 
      \includegraphics[scale=0.4,angle=0]{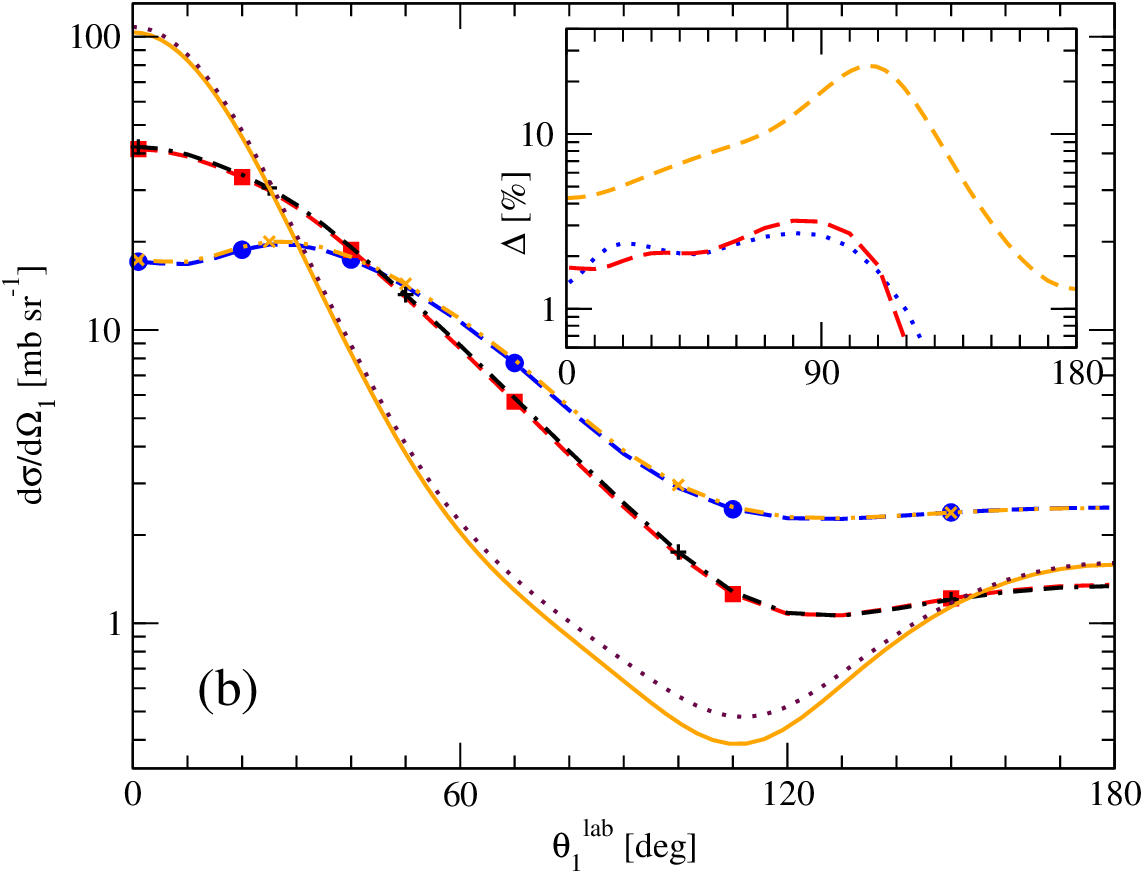}
    \end{center}
    \vspace{1em} 
  \begin{center}
    \includegraphics[scale=0.4,angle=0]{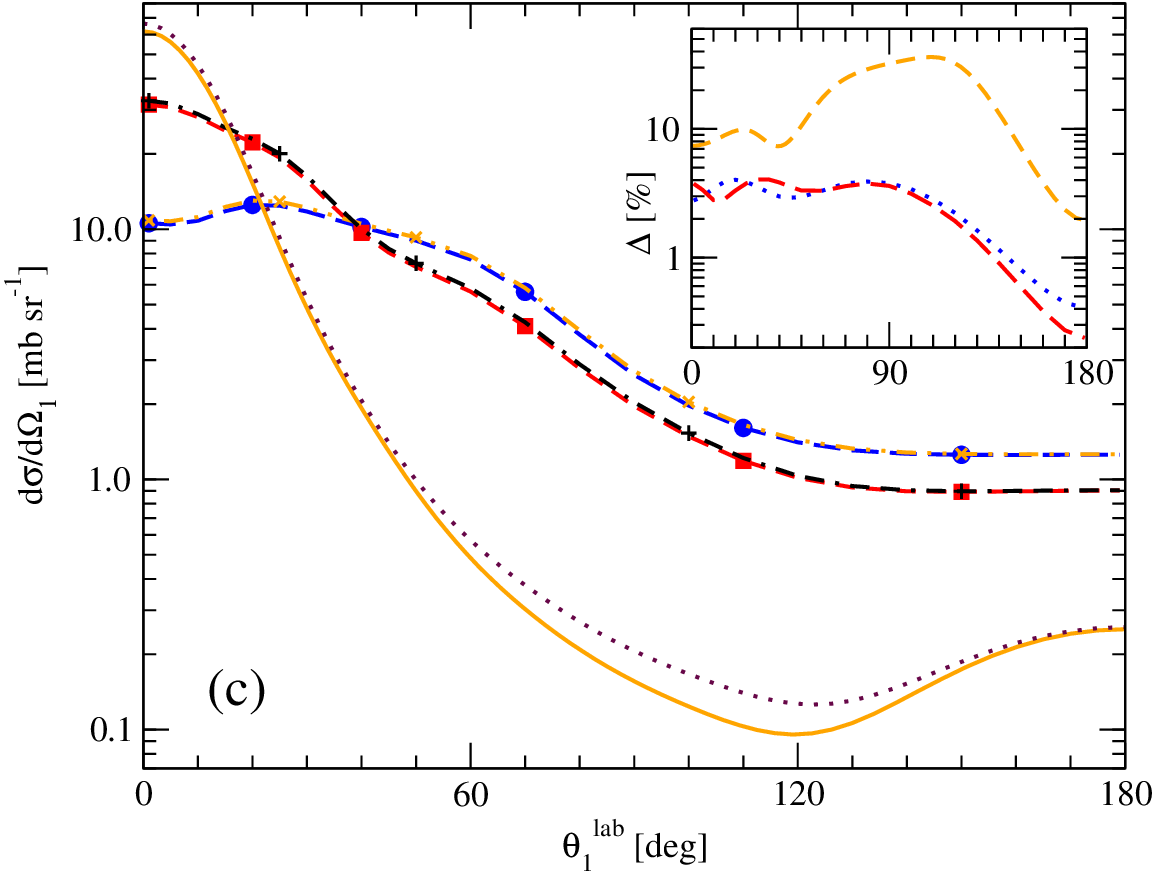}
    \hspace{1em} 
    \includegraphics[scale=0.4,angle=0]{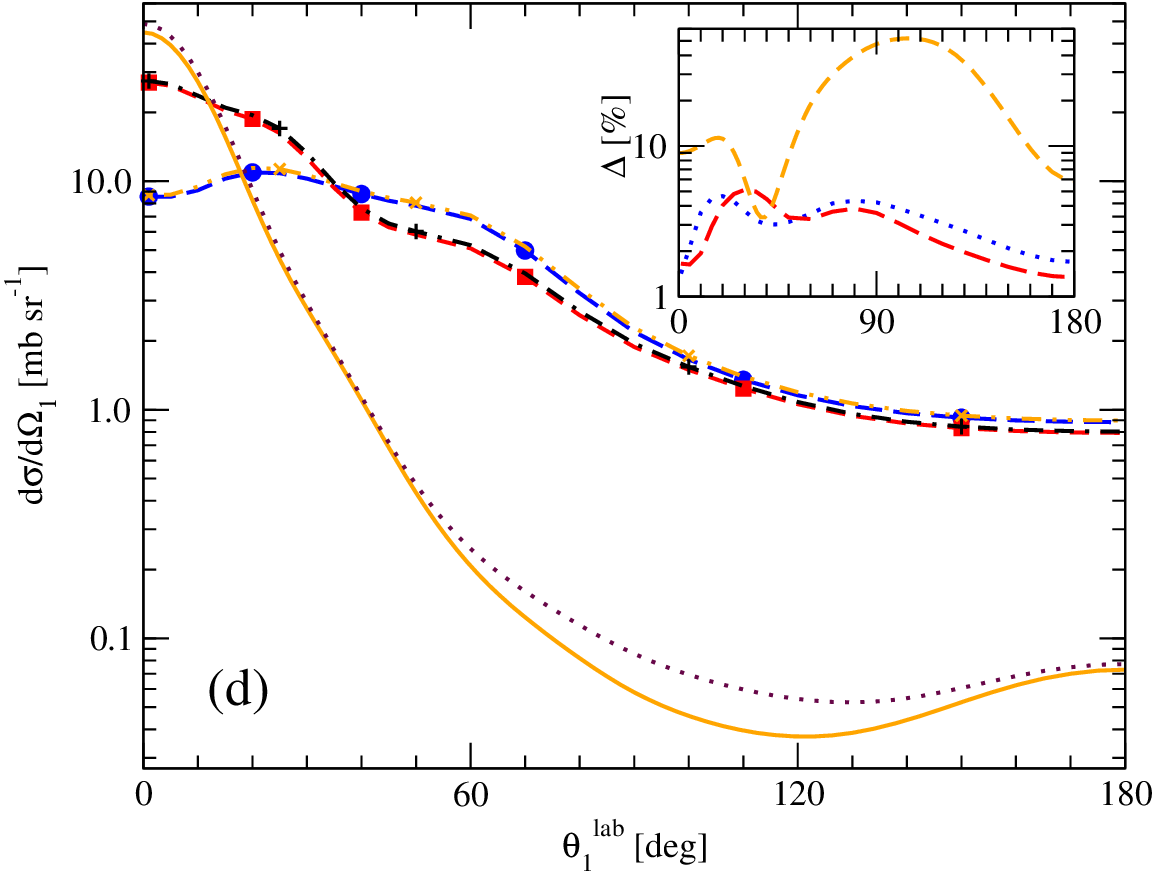}
     \end{center}
\caption{(Color online) 
  The energy spectra from Fig.~\ref{fig2} integrated over the energy of the outgoing
  nucleon shown as a function of the laboratory angle of that nucleon.
  The incoming nuleon kinetic energy in the laboratory system is $E$=14 (a), 70 (b), 135 (c) and 200 (d)~MeV.
  The angular distribution of the outgoing neutron in d(n,n)np reaction is 
  given with the (blue) dashed line, marked with circles when nucleons interact
  with the CD~Bonn potential only. Combining the CD~Bonn with the TM99 3NF gives
  the (orange) dashed-double-dotted line, marked with x-es.
  The angular distribution  of the outgoing
  proton is given by the (red) dashed line, marked with squares,
  (the CD~Bonn only) and the (black) double-dashed-dotted line, marked with pluses,
  (the CD~Bonn + TM99). The laboratory angular distributions for the elastic nd
  scattering is given by the (orange) solid line
  (the CD~Bonn only) and the (maroon) dotted line 
  (te CD~Bonn + TM99). For explanation of inserts see Fig.~\ref{fig2}.
  In inserts shown here the additional
   (orange) short-dashed line represents the magnitude 
  of 3NF effects for elastic nd scattering as defined in caption to
  Fig.~\ref{fig2}.
\label{fig4}}
\end{figure}

To investigate that issue we studied the angular distributions of the incomplete
breakup cross sections 
and compared them to the ones in elastic nd scattering.  
The  spectra at a specific angle integrated over energy of the outgoing nucleon
are shown in Fig.~\ref{fig4}
as functions of the laboratory angle of the
detected nucleon. In this figure also the laboratory 
angular distributions of the cross section
for elastic nd scattering are presented together
with magnitudes of 3NF effects for both elastic scattering and the breakup reaction
shown in inserts of that figure.
The 3NF effects in the breakup, contrary to elastic scattering,
are restricted to the 
 angles below $\approx 120^{\circ}$ and are similar in magnitude for the detected
 neutron or proton. At $70$~MeV effects are $\approx 2-3\%$ up to about
  $\theta_1^{lab} \approx 100^{\circ}$, at
 $135$~MeV $\approx 3-4\%$ up to $\theta_1^{lab} \approx  120^{\circ}$, and
 at $200$~MeV $\approx 4-5\%$ up to
  $\theta_1^{lab} \approx 110^{\circ}$.
With the increasing energy they extend to bigger angles and at $200$~MeV they
reach $\approx 2\%$ for $\theta_1^{lab} \approx 150^{\circ}$. The 3NF effects
 in the angular distributions of the breakup process are significantly smaller than
those in elastic Nd scattering, where at $200$~MeV 
they amount to $\approx 50\%$ in the region of the cross section minimum (see
Fig.~\ref{fig4}). 

\begin{figure}[htbp] 
\includegraphics[scale=0.5]{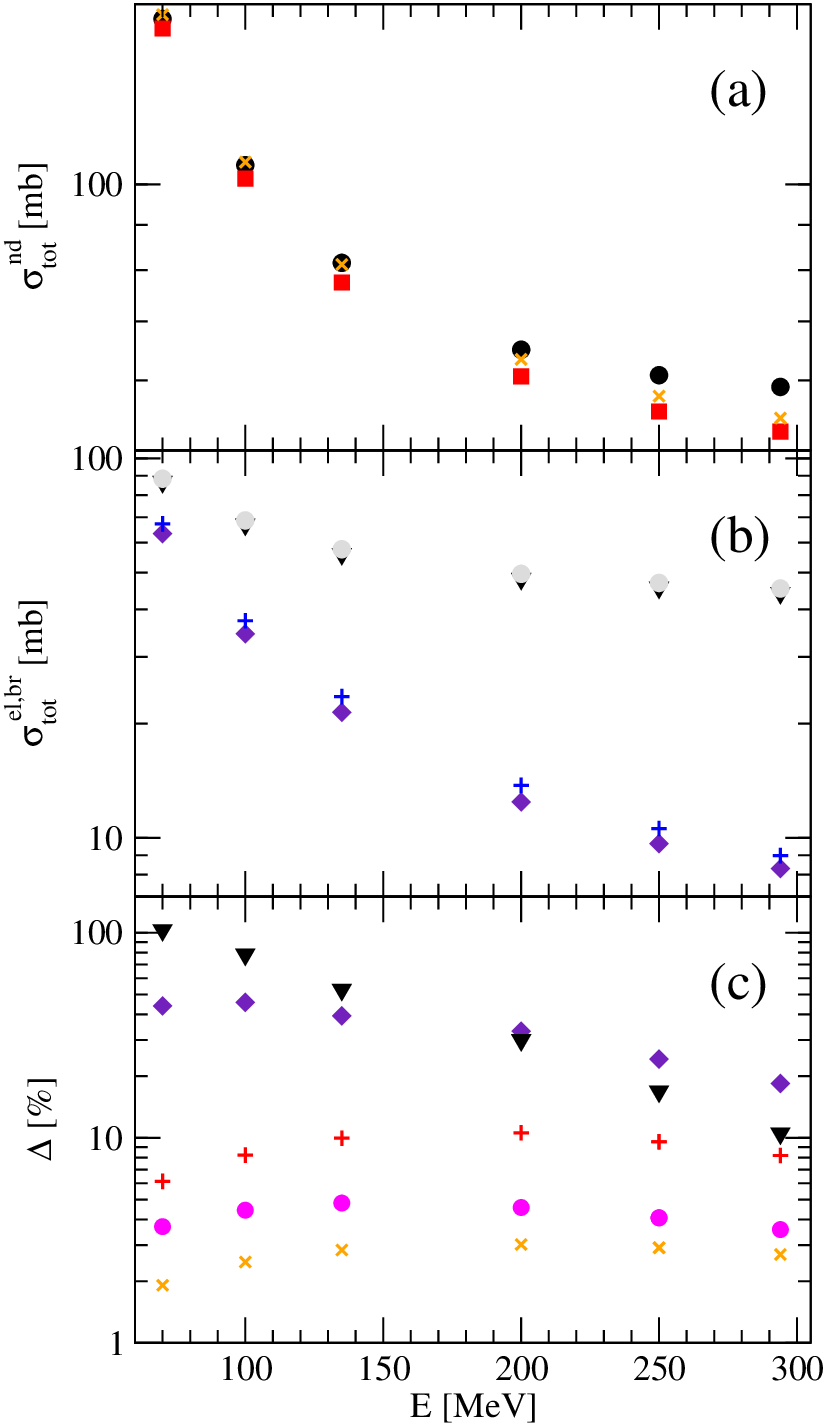}
\caption{(Color online)
  The total cross section for nd scattering (a) and the 
  total nd elastic scattering
  and the total breakup cross sections (b) 
  as functions of the laboratory energy $E$ of the
  incoming neutron. In (a) (black) dots demote experimental total cross section
  data from Ref.~\cite{abfalt98} and (red) squares and (orange) x-es 
  represent the CD Bonn
  and the CD Bonn + TM99 based total nd cross section predictions, respectively.
  In (b) the corresponding predictions for the total elastic scattering
  cross sections  are given by (indigo) diamonds and (blue) pluses, while for
  the total breakup cross section by (black) triangles and (grey) dots.
  In (c) the contributions from the elastic scattering and breakup 
  reactions provided by
  the TM99 3NF to the difference  between 
  the experimental total nd cross section and its 
  prediction by the CD Bonn potential only, defined by
  $\Delta \equiv \frac { \sigma^{tot,NN+3NF}_{el,br} - \sigma^{tot,NN}_{el,br}}
  {\sigma^{tot}_{exp} - \sigma^{tot,NN}} \times 100 \%$, are shown 
  by (indigo) diamonds for elastic scattering and by (black) triangles for
  breakup contribution. The (magenta) dots, (red) pluses, and (orange) x-es
  show the magnitude of the TM99 3NF effects, defined by
   $\Delta \equiv \frac { \sigma^{tot,NN+3NF}_{tot,el,br} - \sigma^{tot,NN}_{tot,el,br}}
  {\sigma^{tot,NN}_{tot,el,br}} \times 100 \%$, for the total nd cross section,
  as well as for the  elastic scattering and breakup total cross sections,
  respectively.
\label{fig5}}
\end{figure}

In Fig.~\ref{fig5}a we show, as functions of the laboratory energy $E$
of the incoming neutron, the total nd
cross section data from Ref.~\cite{abfalt98} together with the
CD~Bonn and the CD~Bonn + TM99 based
theoretical predictions. Similarly to the elastic scattering angular
distributions of the cross sections, 3NF effects start to appear 
in the total nd cross section  at about 60~MeV.
Standard models of 3NF such as the TM99 or the UrbanaIX are able to explain the
difference between the total nd cross section data and theoretical predictions
based on  NN
 potentials  up to
 $\approx 135$~MeV \cite{abfalt98,wittot99}. At higher energies,
 however, they fail to 
reproduce the total cross section data, leaving a significant deviation to data which is
 rapidly growing with the energy, as exemplified in  Fig.~\ref{fig5}a for the CD~Bonn NN potential and
 the TM99 3NF model.

The elastic scattering and breakup total cross sections predicted by the CD~Bonn
 potential alone or combined with the TM99 3NF force  are shown
 in Fig.~\ref{fig5}b.
 Since the angular distributions of the cross sections 
for breakup and  elastic scattering
 are peaked at forward angles (see Fig.~\ref{fig4}), the magnitude of 
 3NF effects as given by the TM99 model for the total nd breakup
 and total elastic scattering cross sections
 can be traced back to the angular distributions of 3NF effects in these
 processes.
 A uniform distribution of 3NF effects in the region of angles with large
 breakup cross sections leads to a magnitude of 3NF effects of 
 $\approx 2-3\%$ for the total nd breakup cross section (see Fig.~\ref{fig5}c).
 The main region of large 3NF
 effects for elastic scattering is located around the minimum of the cross
 section, therefore the magnitude of 3NF effects in the total elastic
 scattering cross section is reduced to $\approx 6-10\%$
 (see Fig.~\ref{fig5}c). 
 For energies above $\approx 70$~MeV the dominating contribution to the total
 cross section comes from the breakup reaction as shown in Fig.~\ref{fig5}b 
(see also Ref.~\cite{kuros_br}), 
which results in the 
 magnitude of 3NF effects in the total nd cross section being
 of the order of $\approx 4\%$ for energies above $70$~MeV. 

It is also interesting to study  contributions to the difference
between the total nd cross section data and  the CD~Bonn potential
prediction which are  induced by the TM99 3NF and which come 
 from elastic scattering and breakup processes separately. 
 They are shown in
 Fig.~\ref{fig5}c by diamonds for the elastic scattering and by triangles for
 the breakup contributions.
 At $200$~MeV the TM99 3NF explains  $\approx 60\%$ of that difference
 with approximately equal contributions from the elastic scattering 
and breakup reactions.
 At $250$~MeV only $\approx 45\%$ is explained with $\approx 25\%$ and
 $\approx 20\%$ contributions from the elastic scattering and breakup
 processes, respectively. 
 At $294$~MeV the explained part reduces further to only $\approx 30\%$
 with contributions of $\approx 20\%$ from the elastic scattering  and
 $\approx 10\%$ from the breakup reaction.   

It is clear that in order to explain the total nd
 cross section data at energies
 around $\approx 200$~MeV a 3NF model containing only long-range 
 2$\pi$-exchange  mechanism  is not sufficient.
 The rapid growth of the unexplained part of the total cross section with increasing energy 
 indicates that the mechanism responsible for it must provide contributions that
 also quickly increase with the energy.
 Among possible mechanisms one could consider  short-range components 
 of the 3NF modelled by the $\pi-\rho$ or $\rho-\rho$ 3NF's \cite{coonpena} 
 exchanges or the corresponding short-range
 components of 3NF as provided in the
 framework of the chiral perturbation
 theory \cite{epel2002,3nf_n3lo_long,3nf_n3lo_short,Piarulli_PRL,Girlanda_short_range}.

\section{Summary and Conclusions}
\label{summary}

We investigated the magnitudes and the distributions 
of 3NF effects in incomplete nd 
breakup  based on  solutions of the 3N Faddeev equation with the 
CD~Bonn potential alone or augmented with the TM99 3NF.
Energy spectra of the outgoing neutron or proton were calculated either 
at a specific laboratory angle of the outgoing nucleon
or by integrating over some angular range.
The spectra at a specific angle (the threefold differential cross sections)
reveal structures caused by dominant contributing kinematically complete
configurations such as FSI or QFS. The 3NF effects start to appear
at $\approx 60$~MeV of the incoming neutron laboratory energy.
At forward angles large 3NF effects are located in the lower parts of the spectra,
shifting to larger outgoing nucleon energies with increasing angle.
Integration of the threefold differential cross section given 
at a specific angle over the outgoing nucleon
energy leads to the angular distributions of the 
breakup cross section for 
single-nucleon detection, which is different from the angular
distribution of
the corresponding elastic scattering cross section. 
In contrast to elastic scattering, where the
 interference between the direct $PT$
 and exchange  $PG_0^{-1}$ terms leads to a characteristic minimum of the
 cross section, the angular
 distribution of the breakup reaction 
is peaked at forward angles and recedes with
 the increasing angle. Additionally, the largest 3NF effects are
 localized being uniformly spread at forward angles, 
 as opposed to elastic scattering, where they are dominant 
 in the region of the minimum of the elastic scattering cross section. 
 The long-ranged 2$\pi$-exchange TM99 3NF is unable to
 explain the angular distributions of the elastic scattering cross section 
as well as the total nd cross section data at higher
 energies.
 The difference between the nd total cross section data and the 
theoretical predictions which include the $2\pi$-exchange TM99 3NF
grows rapidly with the increasing energy of the incoming nucleon, 
which indicates that the short-range components
 of the 3NF 
can be responsible for this discrepancy. Such short-range forces
 could probably provide contributions quickly increasing with the energy
 and dominating at higher energies. 
 It will be interesting to examine if short-range components of the 3NF which
 are  consistently derived in the framework of chiral effective field theory
 are able to describe the total nd cross section data.

\begin{acknowledgments}
This study has been performed within Low Energy Nuclear Physics
International Collaboration (LENPIC) project and 
was  supported by the Polish National Science Center 
 under Grants No. 2016/22/M/ST2/00173 and 2016/21/D/ST2/01120. 
 The numerical calculations were performed on the 
 supercomputer cluster of the JSC, J\"ulich, Germany.
\end{acknowledgments}

\appendix
\section{Elastic scattering transition amplitude - PT term}
\label{app_a}

The contribution to the elastic scattering transition amplitude
 $\langle {\phi}' \vert PT\vert\phi\rangle$ is given by \cite{glo96,book}:
\begin{eqnarray}
  \langle {\phi}' \vert PT \vert \phi \rangle &=&
  \sum_{JM} \sum_{\alpha' l_0 \lambda_0 I_0}
  \langle 1 m_{s_0}' I_0 M- m_{s_0}'|J M \rangle
  \langle \lambda_0 M - m_{s_0}' -\mu' \frac {1} {2} \mu'
  | I_0 M - m_{s_0}' \rangle \cr
  &\times& Y^*_{\lambda_0 M - m_{s_0}' -\mu'}(\hat {\bar{q}}) \int q'^2dq'
  \int_{-1}^{1} dx \varphi_{l_0}(\pi_1) \frac {G_{\alpha_0,\alpha'}(q_0q'x)}
      {\pi_1^{l_0}\pi_2^{l_{\alpha'}}}
      \langle \pi_2 q' \alpha' \vert T \vert \phi \rangle ~,
\label{eqa1}
\end{eqnarray}
where $m_{s_0}'$ and $\mu'$ are spin projections of the outgoing deuteron and
neutron in the final nd state $\phi'$ with relative momentum
$\vec {\bar q}$ ($|\vec {\bar q}|=q_0$). The quantum numbers of the partial
waves for given total  angular momentum $J$ (with projection $M$ on the
z-axis defined by the incoming neutron momentum) and
parity $\pi=(-1)^{l+\lambda}$ of the 3N system
 are given by $\alpha=[(ls)j(\lambda \frac {1} {2})I
(jI)J (t \frac {1} {2} T)]$. The angular momentum $l$ and spin $s$ of the
two-nucleon subsystem are coupled to its total angular momentum $j$ and 
the total isospin is $t=0$ or $t=1$. The spectator nucleon orbital 
angular momentum
$\lambda$ coupled with its spin $\frac {1} {2}$ gives its total angular momentum
$I$. The total isospin of the 3N system $T$ results from coupling of $t$ with
spectator nucleon isospin $\frac {1} {2}$. The momentum $\pi_1$ in the deuteron wave function $\varphi_{l_0}$ is given by
$\pi_1=\sqrt{q'^2+\frac {1} {4} q_0^2 + q'q_0x}$ while  momentum
 $\pi_2=\sqrt{q_0^2+\frac {1} {4} q'^2 + q'q_0x}$. The quantity 
${G_{\alpha_0,\alpha'}(q_0q'x)} / {\pi_1^{l_0}\pi_2^{l_{\alpha'}}}$ comes from the
permutation operator $P$.



\bibliography{apssamp}

\begin{thebibliography}{}\label{sec:TeXbooks}
%
\bibitem{wit98}
H. Wita{\l}a {\it et al.}, 
	Phys. Rev. Lett. {\bf 81}, 1183 (1998).
%
\bibitem{witelas2001} H. Wita{\l}a, W. Gl\"ockle, J. Golak, A. Nogga, 
 H. Kamada,  R. Skibi\'nski and  J. Kuro\'s-\.Zo{\l}nierczuk,
 Phys. Rev. C{\bf 63}, 024007 (2001).
%
\bibitem{Rep.Prog.Phys.75.016301}
N. Kalantar-Nayestanaki {\it et al.}, 
Rep. Prog. Phys. {\bf 75}, 016301 (2012).       
%
\bibitem{abfalt98} W.P. Abfalterer {\it et al.}, 
 Phys. Rev. Lett. {\bf 81}, 57 (1998).
%
\bibitem{wittot99} H.Wita{\l}a, H. Kamada,  A. Nogga, W.Gl\"ockle, Ch. Elster,
D.~H\"uber, 
Phys. Rev. C{\bf 59}, 3035 (1999).
\bibitem{kuros_br} J. Kuro\'s-\.Zo{\l}nierczuk, H. Wita{\l}a, J. Golak,
  H. Kamada, A. Nogga,   R. Skibi\'nski, and W. Gl\"ockle, 
 Phys. Rev. C{\bf 66}, 024003 (2002).
%
\bibitem{TM99} 
S. A. Coon, H.K. Han, Few Body Syst., {\bf 30}, 131 (2001).
%
\bibitem{uIX}
	B.~S.\ Pudliner {\it et al.}, Phys.\ Rev.\ C{\bf 56}, 1720
	(1997).
%
\bibitem{AV18}
R.~B.~Wiringa {\it et al.},
	Phys.\ Rev.\ C {\bf 51}, 38 (1995).
%
\bibitem{cdb} R. Machleidt,
                 Phys. Rev. C{\bf 63}, 024001 (2001).
%
\bibitem{nijm}
V.~G.~J.~Stoks {\it et al.},
	Phys.\ Rev.\ C {\bf 49}, 2950 (1994).
%
\bibitem{witrel1} H. Wita{\l}a, J. Golak, W. Gl\"ockle, H. Kamada,
Phys. Rev. {\bf C71}, 054001 (2005).
%
\bibitem{witrel2} H. Wita{\l}a {\it et al.},
Phys. Rev. C {\bf 77}, 034004 (2008).
%
\bibitem{wit88}  H. Wita{\l}a, T. Cornelius and W. Gl\"ockle, 
  Few-Body Syst. {\bf{3}}, 123 (1988).
%
\bibitem{glo96} W. Gl\"ockle, H. Wita{\l}a, D. H\"uber, H. Kamada, J. Golak, 
 Phys. Rep. {\bf{274}}, 107 (1996).
%
\bibitem{hub97} D. H\"uber, H. Kamada, H. Wita{\l}a, and W. Gl\"ockle,
                Acta Phys. Polon. {\bf B28}, 1677 (1997).
%
\bibitem{book} W. Gl\"ockle, {\it The Quantum Mechanical Few-Body Problem} 
(Springer-Verlag, Berlin, 1983).
%
\bibitem{coonpena} S.A. Coon and M.T. Pe\~na, 
 Phys.\ Rev.\ C {\bf 48}, 2559 (1993).
%
\bibitem{epel2002} E. Epelbaum {\it et al.}, 
 Phys.\ Rev.\ C {\bf 66}, 064001 (2002).
%
\bibitem{3nf_n3lo_long} V. Bernard, E. Epelbaum, H. Krebs, and
  U.-G. Mei{\ss}ner,  Phys. Rev. C {\bf 77}, 064004 (2008).
%
\bibitem{3nf_n3lo_short} V. Bernard, E. Epelbaum, H. Krebs, and
  U.-G. Mei{\ss}ner,  Phys. Rev.  C{\bf 84}, 054001 (2011).

\bibitem{Piarulli_PRL} M. Piarulli {\it et al.}, Phys. Rev. Lett. {\bf 120}, 052503 (2018).

\bibitem{Girlanda_short_range} L. Girlanda, A. Kievsky, M. Viviani, and L. E. Marcucci,
Phys. Rev.  C{\bf 99}, 054003 (2019).

%
\end{thebibliography}

\newpage
\newpage
\newpage

\newpage

\newpage
\newpage
\newpage

\end{document}